\documentclass[fleqn,10pt]{wlscirep}
\usepackage[utf8]{inputenc}
\usepackage[T1]{fontenc}
\usepackage{tcolorbox}

\title{Diverse Misinformation: Impacts of Human Biases on Detection of Deepfakes on Networks}

\author[1,2,*]{Juniper Lovato}
\author[1]{Jonathan St-Onge}
\author[1,3]{Randall Harp}
\author[1]{Gabriela Salazar Lopez}
\author[1]{Sean P. Rogers}
\author[2]{Ijaz Ul Haq}
\author[1,2]{Laurent H\'ebert-Dufresne}
\author[1,2] {Jeremiah Onaolapo}
\affil[1]{Vermont Complex Systems Center, University of Vermont, Burlington, 05405, U.S.A.}
\affil[2]{Department of Computer Science, University of Vermont, Burlington, 05405, U.S.A.}
\affil[3]{Department of Philosophy, University of Vermont, Burlington, 05405, U.S.A.}

\affil[*]{juniper.lovato@uvm.edu}


\begin{abstract}
Social media platforms often assume that users can self-correct against misinformation. However, social media users are not equally susceptible to all misinformation as their biases influence what types of misinformation might thrive and who might be at risk. We call ``diverse misinformation'' the complex relationships between human biases and demographics represented in misinformation. To investigate how users' biases impact their susceptibility and their ability to correct each other, we analyze classification of deepfakes as a type of diverse misinformation. We chose deepfakes as a case study for three reasons: 1) their classification as misinformation is more objective; 2) we can control the demographics of the personas presented; 3) deepfakes are a real-world concern with associated harms that must be better understood. Our paper presents an observational survey (N=2,016) where participants are exposed to videos and asked questions about their attributes, not knowing some might be deepfakes. Our analysis investigates the extent to which different users are duped and which perceived demographics of deepfake personas tend to mislead. We find that accuracy varies by demographics, and participants are generally better at classifying videos that match them. We extrapolate from these results to understand the potential population-level impacts of these biases using a mathematical model of the interplay between diverse misinformation and crowd correction.  Our model suggests that diverse contacts might provide ``herd correction'' where friends can protect each other.  Altogether, human biases and the attributes of misinformation matter greatly, but having a diverse social group may help reduce susceptibility to misinformation.
\end{abstract}

\begin{document}

\flushbottom
\maketitle

\thispagestyle{empty}

\section{Introduction}
\label{section:background}

There is a growing body of scholarly work focused on distributed harm in online social networks. From leaky data, \cite{bagrow2019information} and group security and privacy \cite{lovato2022limits} to hate speech, \cite{garland2022impact} misinformation \cite{chesney2019deep} and detection of computer-generated content. \cite{groh2022deepfake} 
Social media users are not all equally susceptible to these harmful forms of content. Our level of vulnerability depends on our own biases. We define ``diverse misinformation'' as the complex relationships between human biases and demographics represented in misinformation. This paper explores deepfakes as a case study of misinformation to investigate how U.S. social media users' biases influence their susceptibility to misinformation and their ability to correct each other. We choose deepfakes as a critical example of the possible impacts of diverse misinformation for three reasons: 1) their status of being misinformation is binary; they either are a deepfake or not; 2) the perceived demographic attributes of the persona presented in the videos can be characterized by participants; 3) deepfakes are a current real-world concern with associated negative impacts that need to be better understood. Together, this allows us to use deepfakes as a critical case study of diverse misinformation to understand the role individual biases play in disseminating misinformation at scale on social networks and in shaping a population's ability to self-correct.

We present an empirical survey (N=2,016 using a Qualtrics survey panel \cite{boas2020recruiting}) observing what attributes correspond to U.S.-based participants' ability to detect deepfake videos. Survey participants entered the study under the pretense that they would judge the communication styles of video clips. Our observational study is careful not to prime participants at the time of their viewing video clips so we could gauge their ability to view and judge deepfakes when they were not expecting them (not explicitly knowing if a video is fake or not is meant to emulate what they would experience in an online social media platform). Our survey also investigates the relationship between human participants' demographics and their perception of the video person(a)'s features and, ultimately, how this relationship may impact the participant's ability to detect deepfake content. 

Our objective is to evaluate the relationship between classification accuracy and the demographic features of deepfake videos and survey participants. Further analysis of other surveyed attributes will be explored in future work. We also recognize that data used to train models that create deepfakes may introduce algorithmic biases in the quality of the videos themselves, which could introduce additional biases in the participant's ability to guess if the video is a deepfake or not.

The Facebook Deepfake Detection Challenge dataset that was used to create the videos we use in our survey was created to be balanced in diversity in several axes (gender, skin-tone, age). We suspect that if there are algorithmic-level biases in the model used resulting in better deepfakes for personas of specific demographics, we would expect to see poorer accuracy across the board for all viewer types when classifying these videos. We do see that viewer groups' accuracy differs based on different deepfake video groups. However, our focus is on the perception of survey participants towards deepfakes' identity and demographics to capture viewer bias based on their perception rather than the model's bias and classification of the video persona's racial, age, and gender identity. Our goal is to focus on viewers and capture what a viewer would experience in the wild (on a social media platform), where a user would be guessing the identity features of the deepfake and then interrogating if the video was real or not with little to no priming.

\begin{figure}[h!]
     \centering
        \includegraphics[width=0.6\linewidth]{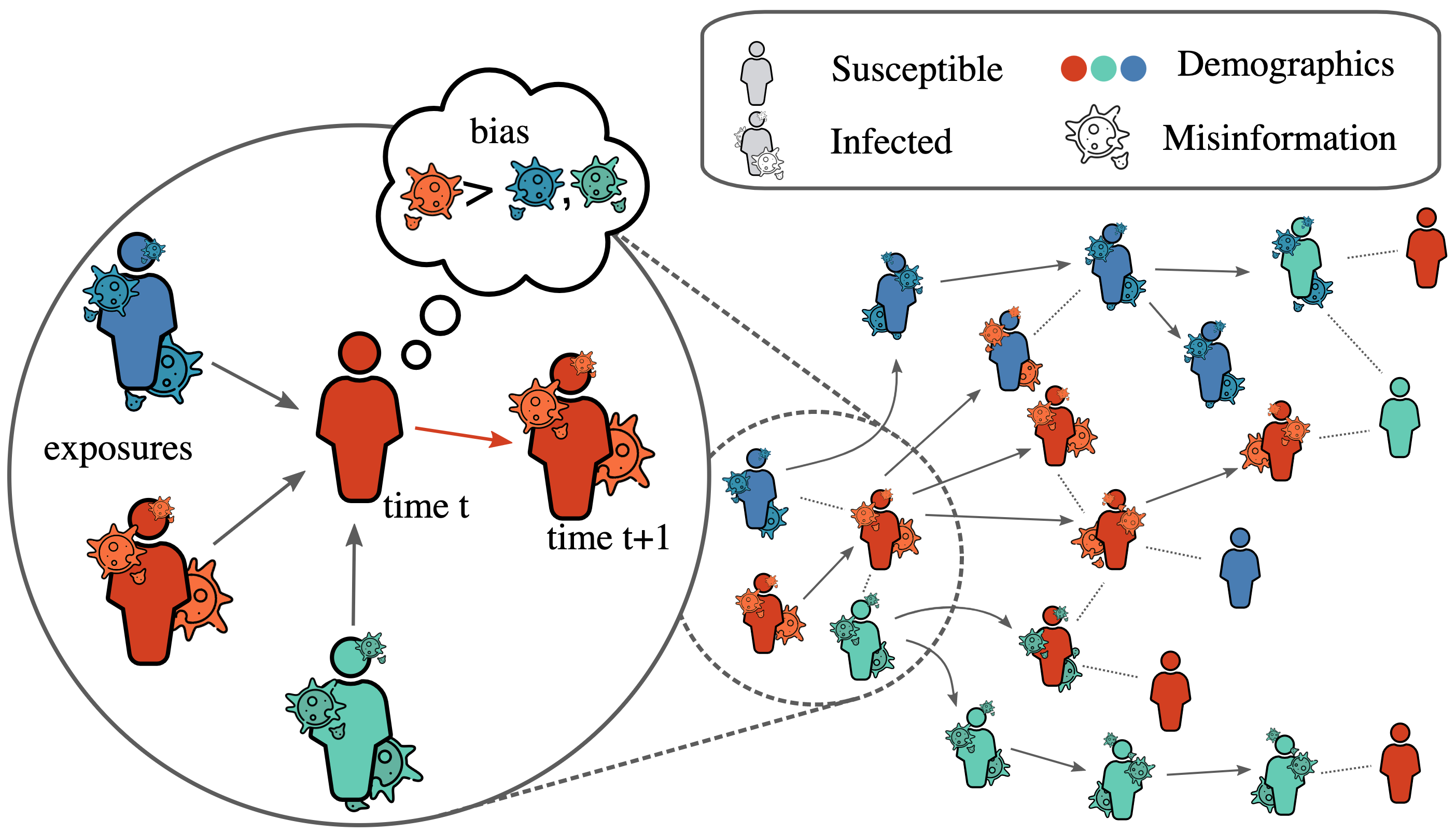}
        \caption{Illustration of the problem considered in this work. Populations are made of individuals with diverse demographic features (e.g., age, gender, race; here represented by colors), and misinformation is likewise made of different elements based on the topics they represent (here shown as pathogens). Through their biases, certain individuals are more susceptible to certain kinds of misinformation. The cartoon represents a situation where misinformation is more successful when it matches an individual's demographic. Red pathogens spread more readily around red users with red neighbors, thereby creating a misinformed echo chamber whose members can not correct each other. In reality, the nature of these biases is still unclear, and so are their impacts on online social networks and on the so-called ``self-correcting crowd''.}
        \label{fig:cartoon}
\end{figure}

This paper adopts a multidisciplinary approach to answer these questions and understand their possible impacts. First, we use a survey analysis to explore individual biases related to deepfake detection. There is abundant research suggesting the demographics of observers and observed parties influence the observer's judgment and sometimes actions toward the observed party. \cite{ebner2020uncovering, lloyd2017black, bond2022racial, klaczynski2020gender, macchi2011age} In an effort to avoid assumptions about any demographic group, we chose four specific biases to analyze vis-\`a-vis deepfakes:
(Question 1) Priming bias: How much does classification accuracy depend on participants being primed about the potential of a video being fake? Our participants are not primed on the meaning of deepfakes and are not told to be explicitly looking for them prior to beginning the survey.  Importantly, we do not explicitly vary the priming of our participants but we compare their accuracy to a previous study with a similar design but primed participants \cite{groh2022deepfake}. Participants are debriefed after the completion of the survey questions and then asked to guess the deepfake status of the videos they watched. More information about our survey methodology and why the study was formulated as a deceptive survey can be seen in section \ref{section:methods}. (Question 2) Prior knowledge: Does accuracy depend on how often the viewer uses social media and whether they have previously heard of deepfakes? Here, we ask participants to evaluate their own knowledge and use their personal assessment to answer this research question. (Question 3) Homophily bias: Are humans better classifiers of video content if the perceived demographic of the video persona matches their own identity? (Question 4) Heterophily bias: Inversely, are humans more accurate if the perceived demographic of the video persona does not match their own? 
We then use results from the survey to develop an idealized mathematical model to theoretically explore population-level dynamics of diverse misinformation on online social networks. Altogether, this allows us to hypothesize the mechanisms and possible impacts of diverse misinformation, as illustrated in Fig.~\ref{fig:cartoon}. 

Our paper is structured as follows. We outline the harms and ethical concerns of diverse misinformation and deepfakes in Section \ref{section:background}. We explore the possible effects through which demographics impact susceptibility to diverse misinformation through our observational study in Section \ref{section:results}. We then investigate the network-level dynamics of diverse misinformation using a mathematical model in Section \ref{section:mathmodel}. We discuss our findings and their implications in Section \ref{section:discussion}. Our full survey methodology can be seen in Section \ref{section:survey}. 

\label{sec:backgroundbais}

Subsequently, these biases impact how social ties are formed and, ultimately, the shape of the social network. For example, in online social networks, homophily often manifests through triadic closures \cite{leskovec2008microscopic} where friends in social networks tend to form new connections that close triangles or triads. Understanding individuals' and groups' biases will help understand the network's structure and dynamics and how information and misinformation spread on the network depending on its level of diversity. For example, depending on the biases and the node-specific diversity of the connections it forms, one may have a system that may be more or less susceptible to widespread dissemination as it would in a Mixed Membership Stochastic Block Model (MMSBM) \cite{airoldi2008mixed}. A Mixed Membership Stochastic Block Model is a Bayesian community detection method that segments communities into blocks but allows community members to mix with other communities. Assumptions in an MMSBM include a list of probabilities that determine the likelihood of communities interacting. We explore these topics in more detail in Section \ref{section:mathmodel}.

Previous work has demonstrated that homophily bias towards content aligned with one's political affiliation can impact one's ability to detect misinformation. \cite{traberg2022birds, calvillo2021personality} Traberg et al. show that political affiliation can impact a person's ability to detect misinformation about political content. \cite{traberg2022birds} They found that viewers misclassified misinformation as being true more often when the source of information aligned with their political affiliation. Political homophily bias, in this case, made them feel as though the source was more credible than it was.

In this paper, we investigate the accuracy of deepfake detection based on multiple homophily biases in age, gender, and race. We also explore other bias types, such as heterophily bias, priming, and prior knowledge bias impacting deepfake detection.

\label{sec:backgroundmisinfo}

Misinformation is information that imitates real information but does not reflect the genuine truth. \cite{lazer2018science} Misinformation has become a widespread societal issue that has drawn considerable recent attention. It circulates physically and virtually on social media sites \cite{watts2021measuring} and interacts with socio-semantic assortativity. In contrast, assortative social clusters will also tend to be semantically homogeneous. \cite{roth_quoting_2022} For instance, misinformation promoting political ideology might spread more easily in social clusters based on shared demographics, further exacerbating political polarization and potentially influencing electoral outcomes \cite{appel2022detection}. This has sparked concerns about the weaponization of manipulated videos for malicious ends, especially in the political realm \cite{appel2022detection}. Those with higher political interests are more likely to share deepfakes inadvertently, and those with lower cognitive ability are also more likely to share deepfakes inadvertently. The relationship between political interest and deepfakes sharing is moderated by network size \cite{ahmed2021inadvertently}.

Motivations vary broadly to explain why people disseminate misinformation, which we refer to as disinformation when specifically intended to deceive. Motivations include 1) purposefully trying to deceive people by seeding distrust in information, 2) believing the information to be accurate and spreading it mistakenly, and 3) spreading misinformation for monetary gain. In this paper, we will primarily focus on deepfakes as misinformation meaning the potential of a deepfake viewer getting duped and sharing a deepfake video. Disinformation is spreading misinformation with the intent to deceive. In this paper, we do not assume that all deepfakes are disinformation since we do not consider the intent of the creator. A deepfake could be made to entertain or showcase technology. We instead focus on deepfakes as misinformation meaning the potential of a deepfake viewer getting duped and sharing a deepfake video, regardless of intent. 

There are many contexts where online misinformation is of concern. Examples include misinformation around political elections and announcements (political harms) \cite{jacobsen2023tensions}; such deepfake videos can, in theory, alter political figures to say just about anything, raising a series of political and civic concerns \cite{jacobsen2023tensions}; misinformation on vaccinations during global pandemics (health-related harms) \cite{chou2018addressing, tasnim2020impact}; false speculation to disrupt economies or speculative markets \cite{kimmel2004rumors}; distrust in news media and journalism (harms to news media) \cite{chesney2019deep, Rini2020}. People are more likely to feel uncertain than to be misled by deepfakes, but this resulting uncertainty, in turn, reduces trust in news on social media \cite{vaccari2020deepfakes}; false information in critical informational periods such as humanitarian or environmental crises \cite{walter2021evaluating}; and propagation of hate speech online \cite{garland2022impact} which spreads harmful false content and stereotypes about groups (harms related to hate speech).

Correction of misinformation: There are currently many ways to try to detect and mitigate the harms of misinformation online. \cite{10.1145/3373464.3373475} On one end of the spectrum are automated detection techniques that focus on the classification of content or on observing anomaly detection in the network structure context of the information or propagation patterns. \cite{starbird2014rumors, sedhai2015hspam14} Conversely, crowd-sourced correction of misinformation leverages other users to reach a consensus or simply estimate the veracity of the content. \cite{10.1145/2998181.2998294, micallef2020role, allen2021scaling} We will look at the latter form of correction in an online social network to investigate the role group correction plays in slowing the dissemination of diverse misinformation at scale. 

Connection with deepfakes: The potential harms of misinformation can be amplified by computer-generated videos used to give fake authority to the information. Imagine, for instance, harmful messages about an epidemic conveyed through the computer-generated persona of a public health official. Unfortunately, deepfake detection remains a challenging problem, and the state-of-the-art techniques currently involve human judgment. \cite{groh2022deepfake}

\label{sec:backgrounddeepfake}

Deepfakes are artificial images or videos in which the persona in the video is generated synthetically. Deepfakes can be seen as false depictions of a person(a) that mimics a person(a) but does not reflect the truth. Deepfakes should not be confused with augmented or distorted video content, such as using color filters or digitally-added stickers in a video.  Creating a deepfake can involve complex methods such as training artificial neural networks known as generative adversarial networks (GANs) on existing media \cite{tolosana2020deepfakes} or simpler techniques such as face mapping. 

Deepfakes are deceptive tools that have gained attention in recent media for their use of celebrity images and their ability to spread misinformation across online social media platforms. \cite{roose2018here}
 
Early deepfakes were easily detectable with the naked eye due to their uncanny visual attributes and movement. \cite{moriuncanny} However, research and technological developments have improved deepfakes, making them more challenging to detect. \cite{chesney2019deep} There are currently several automated deepfake detection methods. \cite{verdoliva_media_2020, jung_deepvision_2020, guera2018deepfake, zotov2020deepfake, Logan280020} However, they are computationally expensive to deploy at scale. As deepfakes become ubiquitous, it will be necessary for the general audience to identify deepfakes independently during gaps between the development of automated techniques or in environments that are not always monitored by automated detection (or are offline). It will also be important to allow human-aided and human-informed deepfake detection in concert with automated detection techniques. 

Several issues currently hinder automated methods: 1) they are computationally expensive; 2) there may be bias in deepfake detection software and training data---credibility assessments, particularly in video content, have been shown to be biased; \cite{haut2021could} 3) As we have seen with many cybersecurity issues, there is a ``cat-and-mouse'' evolution that will leave gaps in detection methodology. \cite{shillair2016cyberseccatandmouse}

Humans may be able to help fill these detection gaps. However, we wonder to what extent human biases impact the efficacy of detecting diverse misinformation. If human-aided deepfake detection becomes a reliable strategy, we need to understand the biases that come with it and what they look like on a large scale and on a network structure. We also posit that insights into human credibility assessments of deepfakes could help develop more lightweight and less computationally expensive automated techniques.

\label{section:harms}

As deepfakes improve in quality, the harms of deepfake videos are coming to light. \cite{greengard2019will} Deepfakes raise several ethical considerations: 1) the evidentiary power of video content in legal frameworks; \cite{chesney2019deep, schwartz1990explaining, fallis2020epistemic} 2) consent and attribution of the individual(s) depicted in deepfake videos; \cite{harris2018deepfakes} 3) bias in deepfake detection software and training data; \cite{haut2021could} 4) degradation of our epistemic environment, i.e., there is a large-scale disagreement between what community members believe to be real or fake, including an increase in misinformation and distrust; \cite{chesney2019deep, Rini2020} and 5) possible intrinsic wrongs of deepfakes. \cite{deruiter:2021} 

It is important to understand who gets duped by these videos and how this impacts people's interaction with any video content. The gap between convincing deepfakes and reliable detection methods could pose harm to democracy, national security, privacy, and legal frameworks. \cite{chesney2019deep} Consequently, additional regulatory and legal frameworks \cite{MDFPA_2018} will need to be adopted to protect citizens from harms associated with deepfakes and uphold the evidentiary power of visual content. False light is a recognized invasion of privacy tort that acknowledges the harms that come when a person has untrue or misleading claims made about them. We suspect that future legal protections against deepfakes might well be grounded in such torts, though establishing these legal protections is not trivial. \cite{schwartz1990explaining, citron:2022}

The ethical implications of deepfake videos can be separated into two main categories: the impacts on our epistemic environment and people's moral relationships and obligations with others and themselves. Consider the epistemic environment, which includes our capacity to take certain representations of the world as true and our taking beliefs and inferences to be appropriately justified. Audio and video are particularly robust and evocative representations of the world. They have long been viewed as possessing more testimonial authority (in the broader, philosophical sense of the phrase) than other representations of the world. This is true in criminal and civil contexts in the United States, where the admissibility of video recordings as evidence in federal trials is specifically singled out in Article X of the Federal Rules of Evidence \cite{FBEvidence_2019} (State courts have their own rules of evidence, but most states similarly have explicit rules that govern the admissibility of video recordings as evidence).  The wide adoption of deepfake technology would strain these rules of evidence; for example, the federal rules of evidence reference examples of handwriting authentication, telephone conversation authentication, and voice authentication but do not explicitly mention video authentication. Furthermore, laws are notorious for lagging behind technological advances, \cite{soloveconceptualizing} which can further complicate and limit how judges and juries can approach the existence of a deepfake video as part of a criminal or civil case.

\label{sec:researchqs}

Our paper asks four primary research questions regarding how human biases impact deepfake detection. (Q1) Priming: How important is it for an observer to know that a video might be fake? (Q2) Prior knowledge: How important is it for an observer to know about deepfakes, and how does social media usage affect accuracy? (Q3-Q4) Homophily and heterophily biases: Are participants more accurate at classifying videos whose persona they perceive to match (homophily) or mismatch (heterophily) their own demographic attributes in age, gender, and race?

To address our four research questions, we designed an IRB-approved survey (N=2,016) using video clips from the Deepfake Detection Challenge (DFDC) Preview Dataset. \cite{DFDC2019Preview, DFDC2020} Our survey participants entered the study under the pretense that they would judge the communication styles of video clips (they were not explicitly looking for deepfake videos in order to emulate the uncertainty they would experience in an online social network).  After the consent process, survey participants were asked to watch two 10-second video clips. After each video, our questionnaire asked participants to rate the pleasantness of particular features (e.g., tone, gaze, likability, content) of the video on a 5-point Likert scale. They were also asked to state their perception of the person in the video by guessing the video persona's gender identity, age, and whether they were white or a person of color.

After viewing both videos and completing the related questionnaire, the participants were then debriefed on the deception of the survey, given an overview of what deepfakes are, and then asked if they thought the videos they just watched were real or fake. After the debrief questions, we collected information on the participants' backgrounds, demographics, and expressions of identity. 

Our project investigates features or pairings of features (of the viewer or the person(a) in the video) that are the most important ones needed to determine an observer's ability to detect deepfake videos and avoid being duped. Conversely, we also ask what pairings of features (of the viewer or the person(a) in the video) are important to determine an observer's likelihood of being duped by a deepfake video.

Our null hypothesis asserts that none of the features or pairing of features we measure in our survey produce biases that show strong evidence of the importance of a user being duped by a deepfake video or being able to detect a deepfake video. We then measure our confidence in rejecting this null hypothesis by measuring a bootstrap credibility interval for a difference in means test between the accuracy of two populations (comparing Matthew's Correlation Coefficient scores).
In all tests, we use 10,000 bootstrap samples and consider a comparison significant (having strong evidence) if the difference is observed in 95\% of samples (i.e., in 9,500 pairs). With this method, our paper aims to better understand how potential social biases affect our ability to detect misinformation. 

\section{Results}
\label{section:results}

Our results can be summarized as follows. (Q1) If not primed, our survey participants are not particularly accurate at detecting deepfakes (accuracy = 51\%, essentially a coin toss). (Q3-Q4) Accuracy varies by some participants' demographics and perceived demographics of video persona. In general, participants were better at classifying videos that they perceived as matching their own demographic.

\begin{figure}[h]
     \centering
        \includegraphics[width=0.6\linewidth]{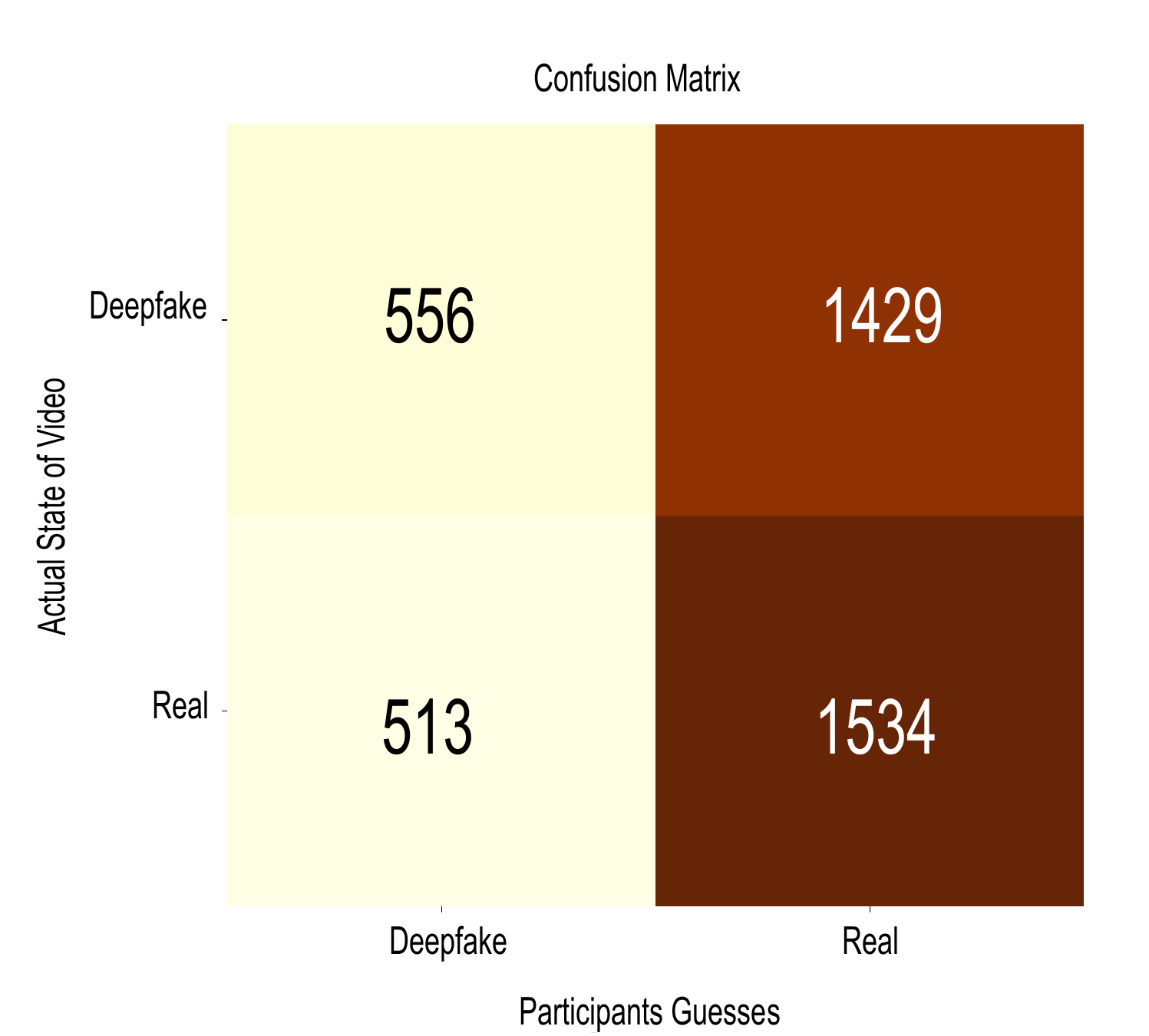}
        \caption{A confusion matrix showing our participant guesses about the state of the videos vs. the real state of the video. Participants in our study watched two videos followed by a questionnaire and a debriefing on deepfakes. They were then asked to guess whether the videos were deepfakes or real. Out of 2,016 participants and 4,032 total videos watched, 1,429 videos duped our participants, meaning they saw a fake video they thought was real. The top right panel shows the participants who were duped by deepfakes. The confusion matrix is defined by the number of true positives in the top left, false negatives in the top right, false positives in the bottom left, and true negatives in the bottom right.}
        \label{fig:guessbytype}
\end{figure}

Our results show that of the 4,032 total videos watched, 49\% were deepfakes, and 1,429 of those successfully duped our survey participants. A confusion matrix showing the True Positive (TP), False Negative (FN), False Positive (FP), and True Negative (TN) rates can be seen in Fig.~\ref{fig:guessbytype}. We also note that the overall accuracy rate (where accuracy = (TP+TN)/(TP+FP+FN+TN))  of our participants was 51\%. This translates to an overall Matthew's Correlation Coefficient (MCC)  score of 0.334 for all participant's guesses vs. actual states of the videos. MCC \cite{matthews1975comparison, boughorbel2017optimal} is a simple binary correlation between the ground truth and the participant's guess. Regardless of the metric, our participants performed barely better than a simple coin flip (credibility 94\%). All summary statistics for our study and all confusion matrices for our primary and secondary demographic groups can be found in Appendix SI2 and Appendix SI3, respectively. Next, we explain our findings in detail.

Q1 Priming bias: Our results suggest that priming bias may play a role in a user's ability to detect deepfakes. Compared with notable prior works, \cite{groh2022deepfake, azur2011multiple, ChengFlock2015} our users were not explicitly told to look for deepfake videos while viewing the video content. Our survey takers participated in a deceptive study where they thought they answered questions about effective communication styles. They were debriefed only after the survey was completed and then asked if they thought the video clips were real or fake. Priming, on the contrary, would mean that when the user watched the two video clips, they would be explicitly looking for deepfakes.

\begin{figure}[h!]
     \centering
        \includegraphics[width=0.5\linewidth]{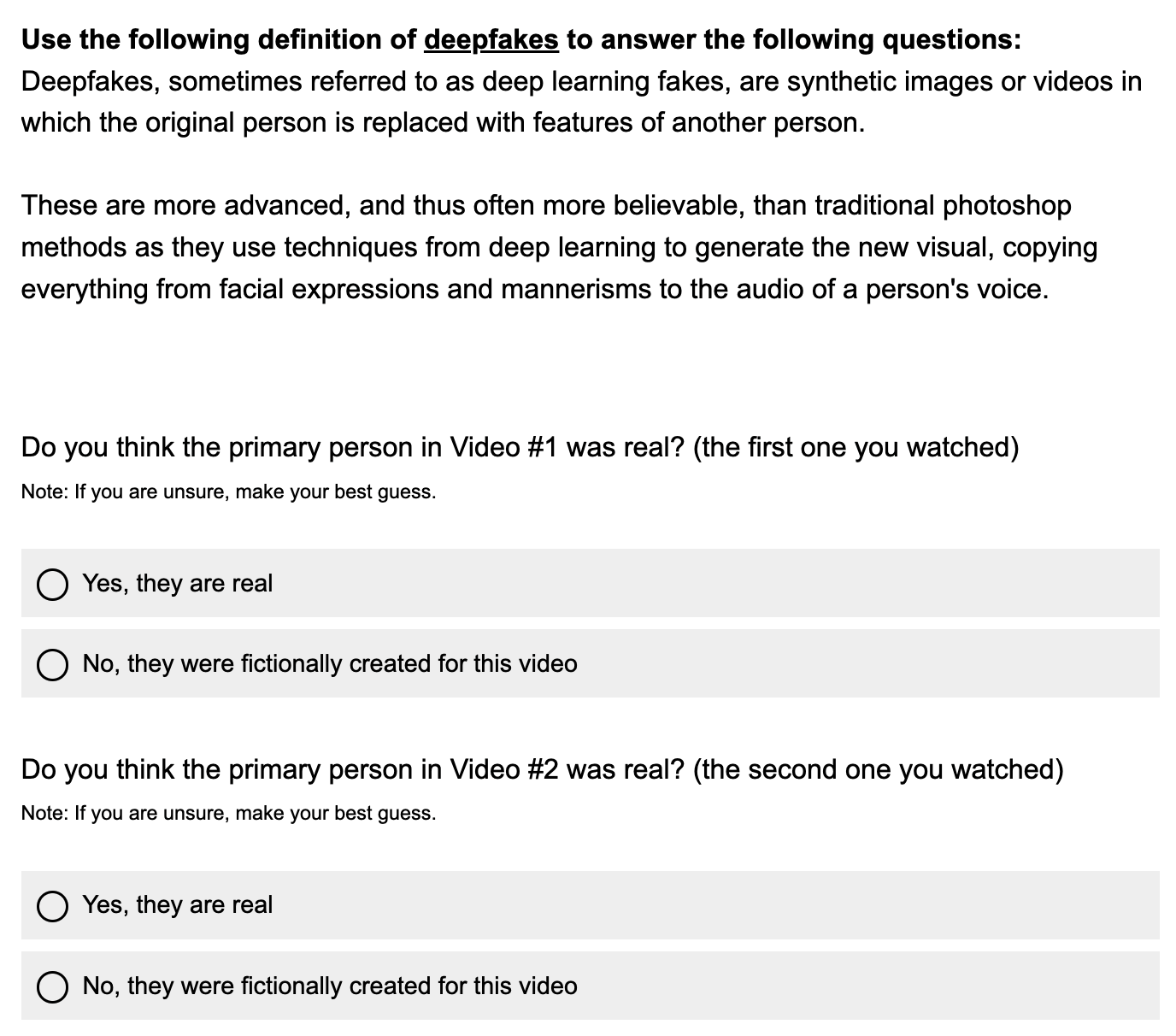}
        \caption{Question where survey participants are asked after the debrief of the survey if they think the videos they watched are real or fake. The performance metric we use to measure participant accuracy is the ratio of the correct guesses to the entire pool of guesses where accuracy = (True Positive (TP) + True Negative (TN))/(True Positive (TP) + False Positive (FP) + False Negative (FN) + True Negative (TN)).}
        \label{fig:accuracy}
\end{figure}

Other works measured primed human deepfake detectors to compare them to machines and humans with machine aid. For example, in a study by Azur et al., \cite{azur2011multiple} humans were deployed as deepfake evaluators. The participants were explicitly asked to view images and look for fake images. Participants in the study were also required to pass a qualification test where they needed to correctly classify 65\% real and fake images to participate in the study \cite{azur2011multiple} fully. In a more recent study by Groh et al., \cite{groh2022deepfake} participants viewed video clips from the Facebook Deepfake Detection Challenge Dataset (DFCD), as in our study. They were asked to explicitly look for deepfake videos and then tested regarding how this compared to machines alone and machines aided by humans. Groh et al. reported an accuracy score of 66\% for primed humans, 73\% for a primed human with a machine helper, and 65\% for the machine alone. In another study, Chen et al. \cite{ChengFlock2015} also showed that hybrid systems that combine crowd-nominated and machine-extracted features outperform humans and machines alone. 

 In comparison, a previous study by Groh et al. ~\cite{groh2022deepfake} uses the same benchmark video data but in their study subjects were informed beforehand and explicitly looked for deepfakes. We compare our participant's accuracy to this study in Table \ref{tab:priming}. The section of the Groh et al. study where they gather human accuracy of deepfakes was conducted through a publicly available website (participant demographics were not gathered for this study). This website collected organic visitors from all over the world, the participants could view deepfakes from the DFCD dataset and guess if they could spot the deepfake or not (the specific question asked ``Can you spot the deepfake video?''), the study participants were asked on a slider how confident they were in their answers as a percentage between 50\%-100\%. In the human detection section of the study, they evaluated the accuracy of 882 individuals (only those who viewed at least 10 pairs of videos, note they did not find evidence that accuracy improves as the participants watch more videos) on 56 pairs of videos from the DFCD dataset. They compare the accuracy rate of participants (66\% for humans alone) in this study with the accuracy of the leading model from the DFDC Kaggle challenge (65\% accuracy for the leading model). In the second part of their experiment, they look at how the leading model (e.g., machine model) can help human accuracy. In this part of the study, after participants (N=9,492) submit their guesses regarding the state of the videos they are given the likelihood from the machine model and then told they can update their scores (resulting in a 73\% accuracy score). 

Our results show that the non-primed participants were only 51\% accurate at detecting if a video was real or fake. One important takeaway from previous studies is that human-machine cooperation provides the best accuracy scores. The previously mentioned prior studies were performed with primed participants. We believe a more realistic reflection of how deepfake encounters would occur ``in the wild'' would be with observers who were not explicitly seeking out deepfakes.  Ecological viewing conditions are important for this type of study. \cite{josephs2023artifact} Future work is needed to investigate how non-primed human deepfake detectors perform when aided by machines.  

\begin{table}[ht]
\centering
\tabcolsep=0.19cm
\caption{Accuracy of deepfake detection}
\begin{tabular}{@{}*{2}{c}@{}}
Type & Accuracy \\ \hline
Non-Primed Human  & 51\% \\ 
Primed Human \cite{groh2022deepfake} & 66\% \\ 
Machine Only \cite{groh2022deepfake}  & 65\% \\ 
Primed Human with Machine Helper \cite{groh2022deepfake} & 73\% \\ \hline
\end{tabular}
\begin{flushleft}
{Accuracy scores of machine deepfake detectors versus primed human deepfake detectors versus non-primed human deepfake detectors. We compare primed and non-primed survey participants and their abilities to detect deepfakes. Our results show that humans who are not primed to find deepfakes reach an accuracy of 51\%. The accuracy scores of our survey participants are 15\% points below those of primed human deepfake detectors from previous work. \cite{groh2022deepfake}}
\end{flushleft}
\label{tab:priming}
\end{table}

Q2 Prior knowledge effect: We also ask if participants are better at detecting a deepfake if they have prior knowledge about deepfakes or more exposure to social media. 

Our results show that there was only weak evidence that prior knowledge or frequent social media usage impacts participants' accuracy. Therefore, we cannot draw any strong conclusions as to the compatibility with our data for this particular question, given that our credibility score for this metric fell below 95\% credibility.

We see that participants who are frequent social media users (i.e., use social media once a week or more) had a higher MCC score (MCC = 0.0396) than those who used social media less frequently (MCC = -0.0110). Participants who knew what a deepfake was before taking the survey (MCC = 0.0790) also had a higher score than those unfamiliar with deepfakes (MCC = 0.0175).  However, in both comparisons, the difference was only deemed to have a weak effect given that bootstrap samples reject the null only with 83\% and 94\% credibility, respectively.

Q3-4 Homophily versus heterophily bias: We then focus on the potential impacts of heterophily and homophily biases on a participant's ability to detect if a video is real or a deepfake.

\begin{figure}
     \centering
        \includegraphics[width=\linewidth]{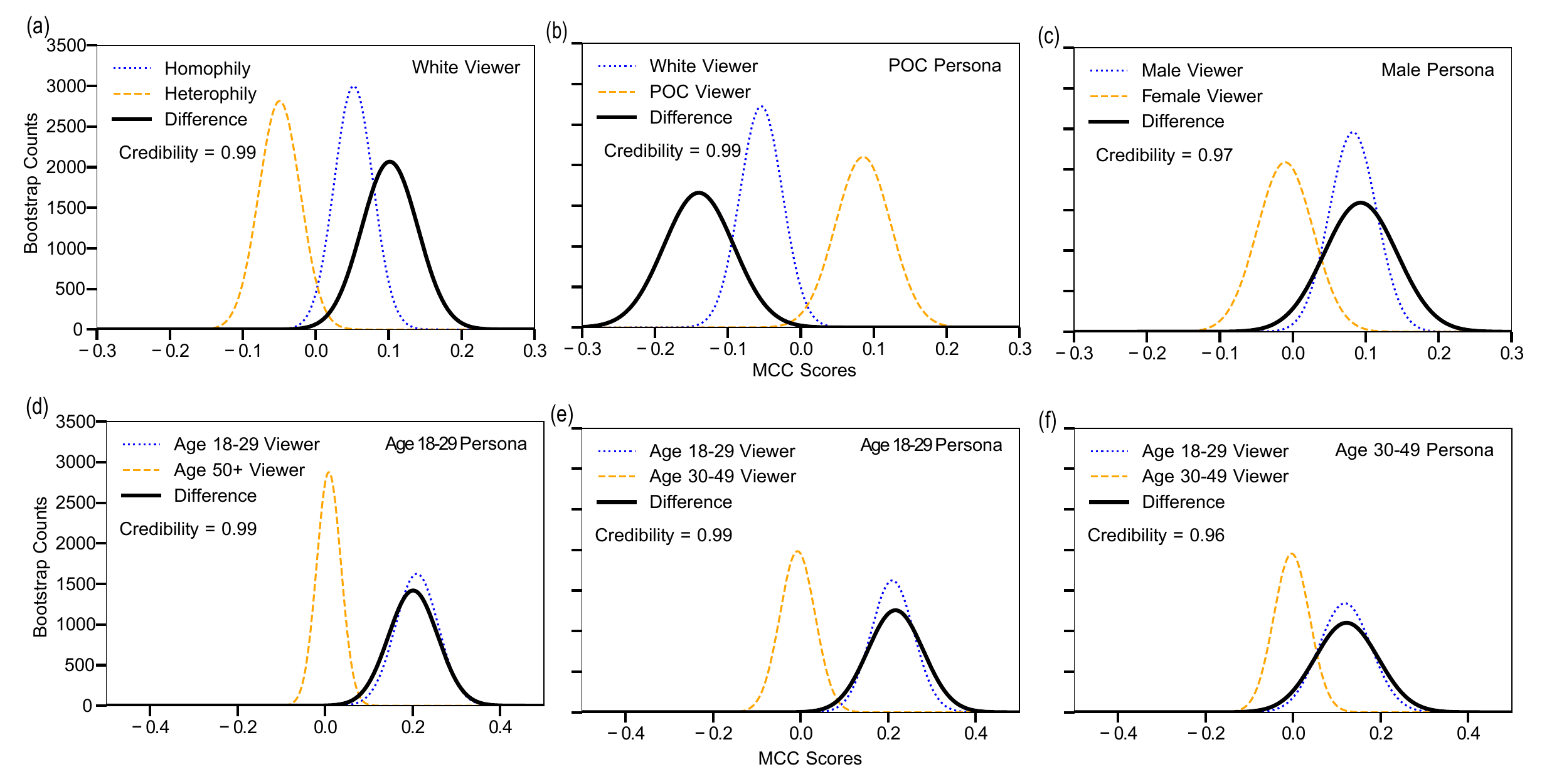}
        \caption{Bootstrap samples from observed confusion matrices to compare MCC scores of user and video feature pairs. Categories that satisfy a threshold of credibility above 95\% are as follows (all bootstrap samples can be seen in the Supplementary Information. (a) White users were found to have a homophily bias and are better at classifying videos of a persona they perceive as white. (b) Consequently, videos of personas of color are more accurately classified by participants of color. (c) Similarly, videos of male personas are better identified by male users. Across multiple age classes, we find that participants aged 18-28 years old are better at identifying videos that match them than older participants (panels (d) and (e)) or even better at classifying videos of persona perceived as 30-49 years old than participants from that same demographic (panel (f)). In addition we reproduce our findings from bootstrapping and conduct a Bayesian logistic regression to explore the effects of matching demographics on the detection accuracy which can be seen in our Supplementary Information}
        \label{fig:bootstrap_all}
\end{figure}

We look at the Matthew's Correlation Coefficients (MCC) for all user groups and compare their guesses on videos that either match their identity (homophily) or do not match their own identity (heterophily). Results of these MCC scores related to homophily and heterophily bias can be seen in Fig.~\ref{fig:bootstrap_all}. 

Our data shows strong evidence that one of our demographic subgroups, namely white participants, was more accurate when guessing the state of video personas that match their own demographic. We test our null hypothesis by comparing the answers given by a certain demographic of participants when looking at videos that match and do not match their identity. In doing so, we only observed evidence of a strong homophily bias for white participants, which can be seen in Table \ref{tab:allsigs}. In that case, the null hypothesis that they are equally accurate on videos of white personas and personas of color falls outside of a 99\% credibility interval, which can be seen in Fig. \ref{fig:bootstrap_all}.

We further break down this potential bias in two dimensions (overall demographic classes of the participants and video persona) in Table ~\ref{tab:allsigs}. We then see more evident results. Here we compare subgroups of our survey participants (e.g., male vs. female viewers, persons of color vs. white viewers, and young vs. old viewers) to see which groups perform better when watching videos of a specific sub-type (e.g., videos of men, videos of women, videos of persons of color, videos of white people, videos of young people, and videos of old people). 

By gender, we find evidence that male participants are more accurate than female participants when watching videos with a male persona. Similarly, by race, we find strong evidence that participants of color are more accurate than white participants when watching videos that feature a persona who is a person of color. Lastly, young participants have the highest accuracy score overall for any of our demographic subgroups.

Of course, these results may be confounded with other factors, such as social media usage, which can be more prominent in one group (e.g., young participants) than another (e.g., older participants). More work needs to be done to understand the mechanisms behind our results.

\begin{table*}[th]
\centering
\tabcolsep=0.19cm
\captionof{table}{Significant differences in accuracy of deepfake detection}
\begin{tabular}{@{}*{5}{c}@{}}
Video/User Demographics  & 	MCC of User   &	{N} & Credibility \\ \hline

White Viewer/Homophilic Videos & 0.0518 & 1372 & 0.99\\ 
White Viewer/Heterophilic Videos  & -0.0498	& 1224 \\
\hline
Male Persona/Male Viewer & 0.0827 & 918 & 0.97\\ 
Male Persona/Female Viewer & 0.0567 & 1188 & \\ 
\hline

POC Persona/POC Viewer & 0.0858 & 708 & 0.99\\ 
POC Persona/White Viewer & -0.0544 & 1143 & \\
\hline

Age 18-29 Persona/Age 18-29 Viewer & 0.1475 & 303 & 0.99\\ 
Age 18-29 Persona/Age 30-49 Viewer & 0.0354 & 264 &\\ 
Age 18-29 Persona/Age 50+ Viewer & -0.0198 & 694 & \\ 

\hline
Age 30-49 Persona/Age 18-29 Viewer & 0.1168 & 282 & 0.96\\ 
Age 30-49 Persona/Age 30-49 Viewer & -0.0037 &  607 &\\

\hline
\end{tabular}
\begin{flushleft}
{Categories that are considered to show strong evidence are ones that satisfy a threshold of credibility above 95\%. Matthew's Correlation Coefficient (MCC) is a correlation measure between a participant's guess about the video being real or fake (0,1) versus the actual state of the video (real 0, fake 1).
We use a bootstrap approach to then test the credibility of a superior accuracy (frequency of bootstrap pairs that produce a superior accuracy). Bootstrap distributions can be seen in Fig.~\ref{fig:bootstrap_all}. Note that ``heterophilic videos'' (row 2) include video personas that the viewers classified as ``maybe POC'' or ``uncertain'', while POC persona (rows 5 and 6) did not.}
\end{flushleft}
\label{tab:allsigs}
\end{table*}

In summary, results that satisfy a threshold of credibility above 95\% (rejecting the null hypothesis with 95\% credibility) on human biases in deepfake detection are as follows.
\begin{itemize}
    \item We find strong evidence that white participants show a homophily bias, meaning they are more accurate at classifying videos of white personas than they are at classifying videos of personas of color.
    \item We find strong evidence that when viewing videos of male personas, male participants in our survey are more accurate than female participants.
    \item We find strong evidence that when viewing videos of personas of color, participants of color are more accurate than white participants. 
    \item We find strong evidence that when viewing videos of young personas, participants between the ages of 18-29 are more accurate than participants above the age of 30; surprisingly, participants aged 18-29 are also more accurate than participants aged 30-49 even when viewing videos of personas aged 30-49.
\end{itemize}

\subsection{Mathematical Model}
\label{section:mathmodel}

In essence, the results shown in Table~\ref{tab:allsigs} illustrate how there is no single demographic class of participants that excels at classifying all demographics of video persona. Different participants can have different weaknesses. For example, a white male participant may be more accurate at classifying white personas than a female participant of color, but the female participant of color may be more accurate on videos of personas of colors. To consider the implications of this simple result, we take inspiration from our findings and formulate an idealized mathematical model of misinformation to better understand how deepfakes spread on social networks with diverse users and misinformation. 

Models of misinformation spread often draw from epidemiological models of infectious diseases. This approach tracks how an item of fake news or a deepfake might spread, like a virus, from one individual to its susceptible network neighbors, duping them such that they can further spread misinformation. \cite{aliberti2017epidemic,jin2013epidemiological,kimura2009efficient,di2013epidemic,shang2015epidemic,scaman2016suppressing,van2022misinformation, weng2013virality} However, unlike infectious diseases, an individual's recovery does not occur on its own through its immune system. Instead, duped individuals require fact-checking or correction from their susceptible neighbors to return to their susceptible state. \cite{bao2013new,zhang2014dynamic,hong2015novel,tambuscio2015fact, xiao2019rumor,zhang2018rumor,kumar2013information,king2022dynamic} In light of these previous modeling studies, it is clear that demographics can affect who gets duped by misinformation and who remains to correct their network neighbors. We therefore integrate these mechanisms with the core finding of our study: Not all classes of individuals are equally susceptible to misinformation.

Our model uses a network with a heterogeneous degree distribution and a structure inspired by the mixed-membership stochastic block model. \cite{airoldi2008mixed} Previous models have shown the importance of community structure for the spread of misinformation~\cite{shang2015epidemic, weng2013virality} and the stylized structure of the mixed-membership stochastic block model captures the known heterogeneity of real networks and its modular structure of echo chambers and bridge nodes with diverse neighborhoods. \cite{red2011comparing} We then track individuals based on their demographics. These abstract classes, such as 1 or 2, could represent a feature such as younger or older social media users. We also track their state, e.g., currently duped by a deepfake video (infectious) or not (susceptible). We also track the demographics of their neighbors to know their role in the network and exposure to other users in different states. 

The resulting model has two critical mechanisms. First, inspired by our survey, individuals get duped by their duped neighbor at a rate $\lambda_i$ dependent on their demographic class $i$. Second, as per previous models and the concept of crowd-sourced approaches to correction of misinformation based on the ``self-correcting crowd'' \cite{10.1145/2998181.2998294, micallef2020role, allen2021scaling}, duped individuals can be corrected by their susceptible neighbors at a fixed rate $\gamma$. The dynamics of the resulting model are tracked using a heterogeneous mean-field approach \cite{pastor2001epidemic} detailed in Box 1 and summarized in Fig.~\ref{fig:model}.

This model has a simple interesting behavior in homogeneous populations and becomes much more realistic once we account for heterogeneity in susceptibility. In a fully homogeneous population, $\lambda_i = \lambda$ $\forall$ $i$, if misinformation can, on average, spread from a first to a second node, it will never stop. The more misinformation spreads, the fewer potential fact-checkers remain. Therefore, misinformation invades the entire population for a correction rate $\gamma$ lower than some critical value $\gamma_c$, whereas misinformation disappears for $\gamma>\gamma_c$.

The invasion threshold for misinformation is shown in Fig.~\ref{fig:model}(a). In heterogeneous populations, where different nodes can feature different susceptibility $\lambda_i$, the discontinuous transition from a misinformation-free to a misinformation-full state is relaxed. Instead, a steady state of misinformation can now be maintained at any level depending on the parameters of misinformation and the demographics of the population. In this regime, we can then further break down the dynamics of the system by looking at the role of duped nodes in the network, as shown in Fig.~\ref{fig:model}(b). The key result here is that very susceptible individuals with a homogeneous assortative neighborhood (e.g., an echo chamber) are at the highest risk of being duped. Conversely, nodes in the same demographic class but with a mixed or more diverse neighborhood are more likely to have resilient susceptible neighbors able to correct them if necessary.

Consider now that diverse misinformation spreads. We assume just two types of misinformation (say young or older personas in two deepfake videos) targeting each of our two demographic classes (say younger and older social media users). We show this thought experiment in Fig.~\ref{fig:model}(c) where we use two complementary types of misinformation: One with $\lambda_1=\lambda_2/2 = 1.0$ and a matching type with $\lambda_2'=\lambda_1'/2 = 1.0$. We run the dynamics of these two types of misinformation independently as we assume they do not directly interact, and, therefore simply combine the possible states of nodes after integrating the dynamical system. For example, the probability that a node of type 1 is duped by both pieces of misinformation would be the product of the probabilities that it is duped by the first and duped by the second. By doing so, we can easily study a model where multiple, diverse pieces of information spread in a diverse network population.

For diverse misinformation in Fig.~\ref{fig:model}(c), we find two connectivity regimes where the role of network structure is critical. For low-degree nodes, a diverse neighborhood means more exposure to diverse misinformation than a homogeneous echo chamber, such that the misinformation that best matches the demographics of a low-degree user is more likely to find them if they have a diverse neighborhood. For high-degree nodes, however, we find the behavior of herd correction: A diverse neighborhood means a diverse set of neighbors that is more likely to contain users who are able correct you if you become misinformed. \cite{walter2021evaluating, bode2015related, vraga2017using}

In the appendix, we analyze the robustness of herd correction to the parameters of the model. We show mathematically that the protection it offers is directly proportional to the homophily in the network (our parameter $Q$). By simulating the dynamics with more parameters, we also find that herd correction is proportional to the degree heterogeneity of the network. As we increase heterogeneity, we increase the strength of the friendship paradox. ``Your friends have more friends than you do,''\cite{feld1991your} which means they get more exposed to misinformation than you do but also that they have more friends capable of correcting them when duped.

\begin{tcolorbox}[title= Box 1: Mathematical model of diverse misinformation and herd correction on social networks, floatplacement=t]
\label{tcolorbox:modelbox}
We wish to explore the potential impacts of our results on the spread of diverse misinformation on social networks. We consider that multiple independent streams of misinformation spread simultaneously; i.e., there are multiple sets of deepfakes, each with its own demographical biases. We also consider that social networks are often very heterogeneous with a skewed distribution of contacts per user and modular with denser connections among users of the same demographics. 

We account for the above using three stylized patterns for the network structure. First, we divide the network into two demographic classes of equal size, simply labeled 1 and 2. Second, we assume a power-law distribution $p_k$ of contacts $k$ per user with $p_k \propto k^{-\alpha}$ regardless of demographics. Third, we use a mixed-membership stochastic block model to generate the network structure: Half of the nodes of each demographic always interact following their demographics, and half act as bridge nodes connecting randomly. The probability that a contact falls within a single demographic class is proportional to $Q$, while contacts across classes occur proportionally to $1-Q$; with $Q>0.5$ for modular structure.

According to the above, we can write the fraction of nodes $p^1_{k,\ell}$ which are of demographic class 1 with $k$ contacts of class 1 and $\ell$ contacts of class 2:
\begin{equation}
    p^1_{k,\ell} \propto \frac{1}{2}(k+\ell)^{-\alpha}\left[\frac{1}{2}\binom{k+\ell}{k}Q^k(1-Q)^\ell + \frac{1}{2}\binom{k+\ell}{k}(1/2)^{k+\ell}\right] \; .
\end{equation}

We define a simple dynamical process where individuals are exposed to misinformation through each of their duped network neighbors, and themselves get duped at a rate $\lambda_i$ based on their demographic class $i$. Non-duped neighbors can then correct their duped neighbors at a rate $\gamma$, \cite{10.1145/2998181.2998294, micallef2020role, allen2021scaling} e.g., we assume that your network neighbors can fact-check something you diffuse online and potentially correct your opinion. The fraction of individuals of a certain type $(i,k,\ell)$ that are duped, $D^i_{k,\ell}$, can be followed in time using a set of ordinary differential equations:
\begin{equation}
    \frac{d}{dt} D^i_{k,\ell} = \lambda_i\left(p^i_{k,\ell}-D^i_{k,\ell}\right)\left(k\theta_{i,1}+\ell\theta_{i,2}\right)-\gamma D^i_{k,\ell}\left(k\phi_{i,1}+\ell\phi_{i,2}\right) \; .
\end{equation}
where $\theta_{i,j}$ and $\phi_{i,j}$ represent the probabilities that a connection from an individual of demographic $i$ to an individual of demographic $j$ connects to a duped or non-duped individual, respectively. They can be calculated, for example, as
\begin{equation}
    \theta_{1,2} = \sum_{k,l} kD^2_{k,\ell} \bigg/ \sum_{k',l'} k'p^2_{k',\ell'} \quad \textrm{or} \quad \phi_{2,1} = \sum_{k,l} \ell \left(p^1_{k,\ell}-D^1_{k,\ell}\right) \bigg/ \sum_{k',l'} \ell'p^1_{k',\ell'} \; .
\end{equation}
These quantities close the system of equations and allow us to simulate a relatively simple model that manages to capture the heterogeneity ($\alpha$) and community structure ($Q$) of social networks, as well as demographic-specific susceptibility to misinformation ($\{\lambda_i\}$) and fact-checking among the population ($\gamma$). Our results are summarized in Fig. \ref{fig:model} and further analyzed in Appendix SI4.

    \begin{minipage}[t]{\linewidth}
    \centering
    \vspace*{0pt}
        \includegraphics[width=\linewidth]{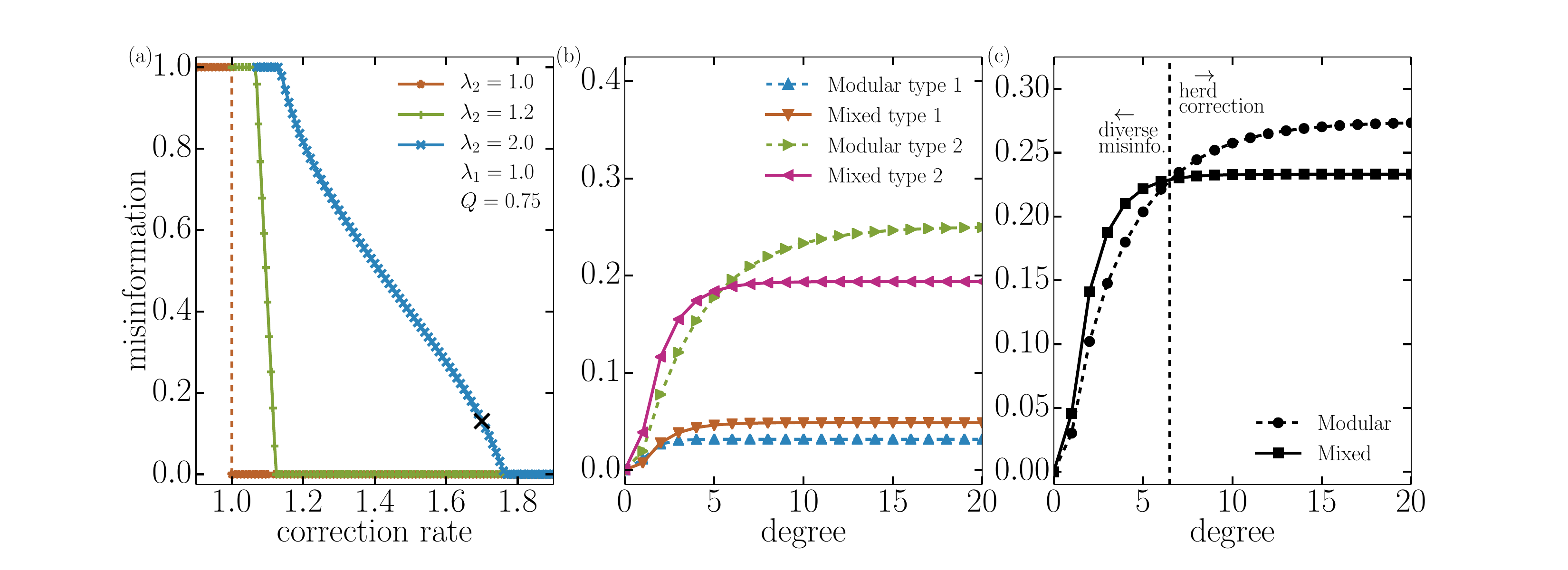}
        \captionof{figure}{Spread of diverse deepfakes with heterogeneous transmission rates $\lambda_i$ across demographic types 1 and 2 (in-group density is set to $Q=0.75$, degree heterogeneity to $\alpha = 3$). Other parameters are given in the figure, with panels (b) and (c) using the correction rate highlighted in (a) at 1.7. Panel (c) shows how high degree nodes can be protected if they have a diverse set of neighbors.}\label{fig:model}
    \end{minipage}
\end{tcolorbox}

Our stylized model is meant to show how one can introduce biases in simple mathematical models of diverse misinformation.
A first-order effect is that individuals with increased susceptibility should be preferentially duped, but this effect exists only if misinformation can spread (above a certain contagion threshold) but not saturate the population (below certain transmissibility such that the heterogeneity has impact). 
A second-order effect is that individuals with a diverse neighborhood are also more likely to have friends who can correct them should they be duped by misinformation.

Future modeling efforts should also consider the possible interactions between different kinds of misinformation. \cite{chang2018co} These can be synergistic, \cite{hebert2015complex} parasitic, \cite{hebert2020spread} or antagonistic; \cite{fu2017dueling} which all provide rich dynamical behaviors. Other possible mechanisms to consider are the adaptive feedback loops that facilitate the spread of misinformation in online social networks. \cite{tornberg2018echo}

\section{Discussion}
\label{section:discussion}

Understanding the structure and dynamics of misinformation is important as it can bring a great amount of societal harm. Misinformation has negatively impacted the ability to disseminate important information during critical elections, humanitarian crises, global unrest, and global pandemics. More importantly, misinformation degrades our epistemic environment, particularly regarding distrust of truths. It is necessary to understand who is susceptible to misinformation and how it spreads on social networks to mitigate its harm and propose meaningful interventions. Further, as deepfakes deceive viewers at greater rates, it becomes increasingly critical to understand who gets duped by this form of misinformation and how our biases and social circle impact our interaction with video content at scale. We hope this work will contribute to the critical literature on human biases and help to better understand their interplay with machine-generated content. 

The overarching takeaways of our results can be summarized as follows. If not primed, humans are not particularly accurate at detecting deepfakes. Accuracy varies by demographics, but humans are generally better at classifying videos that match them. These results appear consistent with findings of the own-race bias (ORB) phenomenon,\cite{meissner2001thirty} where overall, we see that participants are better at detecting videos that match their own attributes. Consistent with ORB research, \cite{anthony1992cross} our study results also show that white participants display a greater accuracy when presented with videos of white personas. We also see strong evidence that persons of color are more accurate than white participants when viewing deepfakes of personas of color and more accurate overall than white participants (see Supplementary Information). Our study adds several extra dimensions of demographic analysis by using gender and age. We see strong evidence that male participants are better at detecting videos of male personas than female viewers. With age, we see strong evidence that when viewing videos of young personas, participants between the ages of 18-29 are more accurate than participants above the age of 30; surprisingly, participants aged 18-29 are also more accurate than participants aged 30-49 even when viewing videos of personas aged 30-49. Combining these results, more work needs to be done to understand better how interventions such as education about deepfakes, cross-demographic experiences and exposure, and exposure to the technology impact a user's ability to detect deepfakes. 

In this observational study, we also explored the potential impacts of these results in a simple mathematical model and extrapolated from our survey to hypothesize that a diverse set of contacts might provide ``herd correction'' where friends can correct each other's blind spots. Friends with different biases can better correct each other when duped. This modeling result is a generalization of the self-correcting crowd approach used in the correction of misinformation. \cite{10.1145/2998181.2998294} 

In future work, we hope to investigate how non-primed human deepfake detectors perform when aided by machines. We want to investigate the mechanisms behind why some human viewers are better at guessing the state of videos that match their own identity. For example, do viewers have a homophily bias because they are more accustomed to images that match their own, or do they simply favor these images? We also would like to empirically investigate our survey via a more robust randomized controlled experiment and model results on real-world social networks with different levels of diversity to measure the spread of diverse misinformation in the wild. Consequently, we would be interested in testing possible educational or other intervention strategies to mitigate adversarial misinformation campaigns. Our simple observational study is a step towards understanding social biases' role and potential impacts in an emerging societal problem with many multilevel interdependencies.

\section{Methods}
\small

\label{section:survey}

\subsection{Survey Methodology} We first ran a pilot stage of our observational study. We conducted a simple convenience sample of 100 participants (aged 18+) to observe the efficacy of our survey. We then ran phase 1 (April-May 2022) of the full survey using a Qualtrics survey panel of 1,000 participants who matched the demographic distribution of U.S. social media users. We then ran phase 2 (September 2022) of the full survey, again using Qualtrics and the same sampling methodology. The resulting full study from phases 1 and 2 is a 2,016-participant sample.
 
Towards ensuring that our experiment reflects the real-world context as closely as possible, survey participants did not know before the start of the survey that the videos could potentially be deepfakes. The survey was framed for participants as a study about different communication styles and techniques that help make video content credible. Participants were told that we were trying to understand how aspects of public speaking, such as tone of voice, facial expressions, and body language, contribute to the effectiveness and credibility of a speaker. The survey's deceptiveness allowed us to ask questions about speaker attributes, likeability, and agreeableness naturally without priming the participants to look specifically for deepfakes. \cite{barrera2012much} We chose to make our survey deceptive not to prime the participants but also because this more closely replicates the deceptiveness that a social media user would encounter in the real world. Furthermore, Br{\"o}der~\cite{broder1998deception} argues that ``in studies of cognitive illusions (e.g., hindsight bias or misleading postevent information effect), it is a necessity to conceal the true nature of the experiment.'' We posit that our study clearly involves cognitive illusions, specifically in the form of deepfakes, and as such deception is an important tool.

We designed our survey using video clips from the Deepfake Detection Challenge (DFDC) Preview Dataset. \cite{DFDC2019Preview, DFDC2020} In our survey, we ask the participants to view two random video clips, which are approximately 10 seconds in length each. Each video clip may be viewed unlimited times before reading the questions but not again after moving to the questions. The information necessary to answer these questions relies solely on the previously shown video clip. A link to the full survey and survey questions is available in Appendix 1.

After viewing both videos, the participants are then asked to complete a related questionnaire about the communication styles and techniques of the videos. The questions ask about attributes of the video, such as pose, tone, and style and are asked to rate them on a Likert scale from very pleasant to very unpleasant. We also asked them to rate their agreement with the video content and credibility. We also ask participants to identify the perceived gender expression of the person(a) in the video, to identify what age group they belong to, and to ask if they perceive the person in the video to be a person of color or not.  

In line with best practices in ethical research~\cite{greene2022best,boynton2013exploring}, we debriefed the participants following the viewing of both videos and completion of the questionnaire on communication style and perceived demographics. The participants are debriefed on the deception of the survey, given a short explanation of deepfake technology, and then asked if they think the videos were real or fake.

Lastly, we collect demographic information on the survey participants' backgrounds and expressions of identity. We also ask participants how knowledgeable they already were on deepfakes, how often they use social media, and their political and religious affiliations. We also asked participants if they knew that the survey was about deepfakes before taking the survey (survey participants who were primed were subsequently dropped from the analysis). 

\subsection*{Sampling}Survey responses from 2,016 participants were collected through Qualtrics, an IRB-approved research panel provider, via traditional, actively managed, double-opt-in research panels. \cite{boas2020recruiting} Qualtrics’ participants for this study were randomly selected stratified samples from the Qualtrics panel membership pool that represents the average social media user in the U.S. \cite{2021pewsocial} Our survey respondents represent the following categories and demographic breakdown in Table \ref{tab:descdfdc2}.

\label{section:data}
\label{section:seconddata}

\subsection{Secondary Data}*For this project, we use the publicly available Facebook AI Research Deepfake Detection Challenge (DFDC) Preview Dataset (N = 5,000 video clips). \cite{DFDC2019Preview,DFDC2020} For our purposes, we filtered out all videos from the dataset that featured more than one person(a). The video clips may be deepfake or real; see Table \ref{tab:descdfdc}. Additionally, some of the videos have been purposefully altered in several ways. Here is the list of augmenters and distractors:
\begin{itemize}
    \item Augmenters: Frame-rate change, Quality level, Audio removal, Introduction of audio noise, Brightness or contrast level, Saturation, Resolution, Blur, Rotation, Horizontal flip.
    \item Distractors: Dog filter,  Flower filter, Introduction of overlaid images, shapes, or dots, Introduction of additional faces,  Introduction of text.
\end{itemize}

A video's deepfake status (deepfake or not) was not revealed to the respondents during or after the survey. Many augmenters and distractors were noticeable to the respondents but were not specifically revealed.\\

\begin{figure}[]
     \centering
        \includegraphics[width=0.6\linewidth]{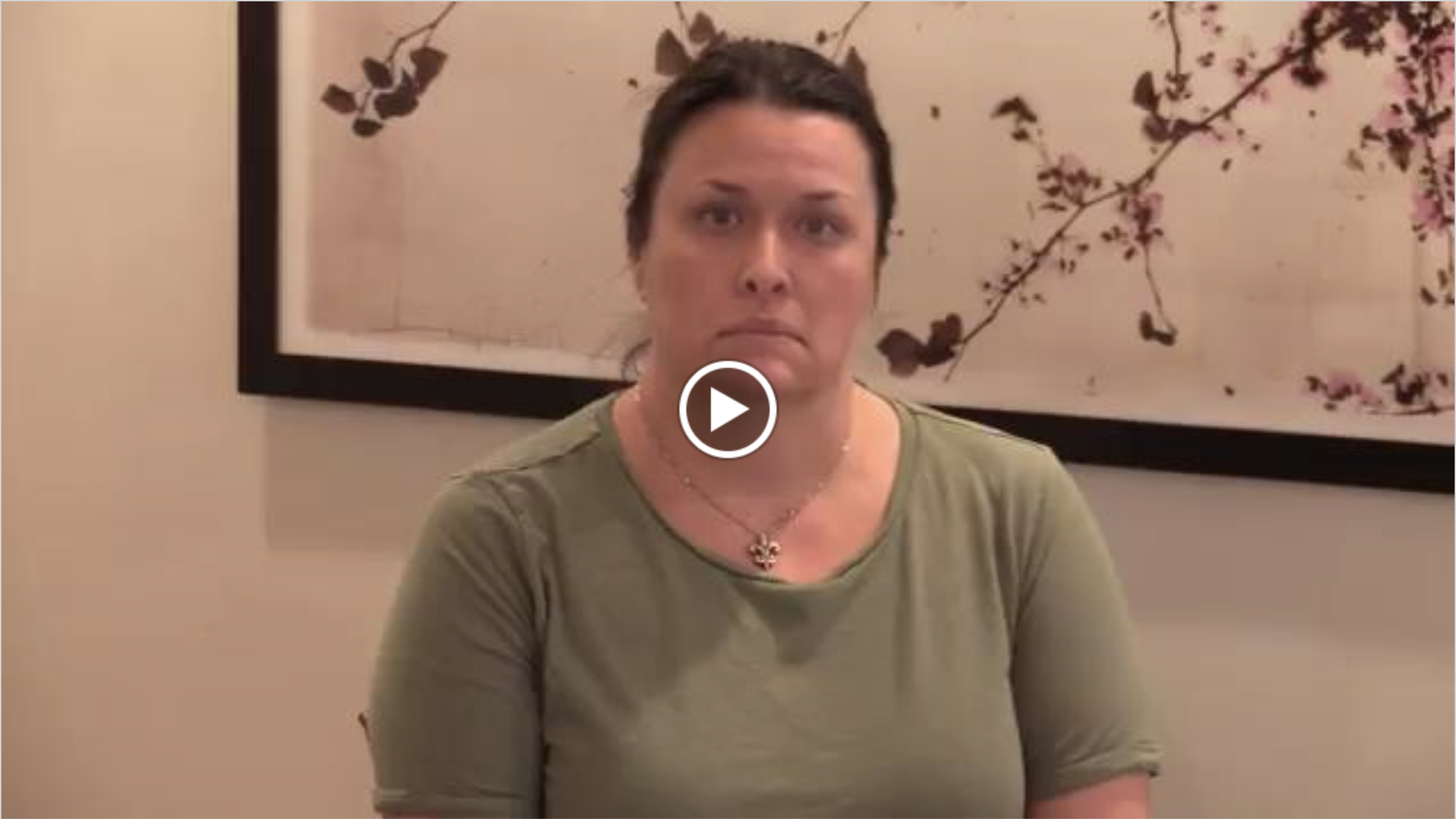}
        \caption{Example video clip from the Facebook Deepfake Detection Challenge (DFDC) dataset. The person depicted is fake.}
        \label{fig:videoimage}
\end{figure}

\begin{table}[ht]
\centering
\caption{Descriptive statistics of video data (N=5,000).}
\tabcolsep=0.19cm
\begin{tabular}{@{}*{4}{c}@{}}
Video & Binary & Number & Percent \\ \hline
Real & 0 & 2,500 & 50\% \\ 
Deepfake & 1 & 2,500 & 50\% \\ \hline
\end{tabular}
\label{tab:descdfdc}
\end{table}

\begin{table}[ht]
\centering
\caption{Descriptive demographics of survey participants (N=2,016).}
\tabcolsep=0.19cm
\begin{tabular}{@{}*{3}{c}@{}}
Type & Sub-Group & \% of Sample \\ \hline
Gender & Female & 45\% \\
Gender & Male &  55\% \\
Gender & Non-Binary & 0.5\% \\
Ages & 18-29 &  17\% \\
Ages&  30-49 &  27\% \\ 
Ages & 50-64 &  29\% \\
Ages &  65+ &  27\% \\
Demographic &  Non-Hispanic White &  67\% \\
Demographic &  Non-Hispanic Black &  10\% \\
Demographic &  Hispanic &  14\% \\
Demographic &  Other &  9\% \\ \hline
\end{tabular}
\label{tab:descdfdc2}
\end{table}

\label{section:ogdata}

\subsection{Original Data} We transformed all survey response variables of interest into numerical form to analyze our survey results. All Likert survey questions were converted from `Very unpleasant,' `Unpleasant,' `Neutral,' `Pleasant,' and `Very pleasant' to an ordinal scale of 1,2,3,4,5. 

Participants selected education levels from `Some high school,' `High school diploma or equivalent,' `Some college,' Associate’s degree (e.g., A.A., A.E., A.F.A., AS, A.S.N.),' `Vocational training,' `Bachelor’s degree (e.g., B.A., BBA BFA, BS),' `Some postgraduate work,' `Master’s degree (e.g., M.A., M.B.A., M.F.A., MS, M.S.W.)),' `Specialist degree (e.g., EdS),' `Applied or professional doctorate degree (e.g., M.D., D.D.C., D.D.S., J.D., PharmD), `Doctorate degree (e.g., EdD, Ph.D.)' was transformed to an ordinal scale of 1-11 respectively. 

Participants selected income levels from `Less than \$30,000', `\$30,000-\$49,999', `\$50,000-\$74,999',  `\$75,000+' were transformed to an ordinal scale of 1-4 respectively.

Participants selected their social media usage levels from `I do not use social media,' `I use social media but less than once a month,' `Once a month,' `A few times a month,' `Once a week,' `A few times a week,' `Once a day,' `More than once a day' were transformed to an ordinal scale of 1-8 respectively. Variables were split into the category of frequent social media users 5-8 and infrequent social media users 1-4. We combined the ordinal scales into two categories in order to reduce the dimensionality of our data. 

Participants selected their knowledge of deepfake from `I did not know what a deepfake was,' `I somewhat knew what a deepfake was,' `I knew what a deepfake was,' `I consider myself knowledgeable about deepfakes' was transformed to an ordinal scale of 1-4 respectively. Variables were split into users who are knowledgeable about deepfakes 3-4 and users who are not knowledgeable about deepfakes 1-2. We combined the ordinal scales into two categories in order to reduce the dimensionality of our data. 

All nominal and categorical variables were transformed into binary variables. Categorical variables (some survey questions included write-in answers) were combined into coarser-grained categories for analysis, such as participant racial/ethnic identity (transformed to Person of Color or White), U.S. state of residence (transformed to U.S. regions), employment (transformed to occupational sectors), religious affiliation (transformed into religious affiliations), and political affiliation (transformed to major political affiliations). 

We allowed survey participants to identify their gender identity, the results of which were largely binary. Unfortunately, our sample was insufficient to perform meaningful analysis on a larger non-binary gender identity spectrum. Primary variables with an N under 30 were dropped, meaning the participant's responses were not included in the analysis (this was only applicable for non-binary gender responses where N=13). Our survey participants were given two video clips to view and critique; in our analysis, we decided to analyze the first or second video in the same way. 

\label{section:methods}

\subsection{Analytical Methods} We use Matthews Correlation Coefficient to understand the relationship between the participant's guesses on the status of the video (fake or real) and the actual state of the video (fake or real), we ran a Matthews Correlation Coefficient (MCC) \cite{matthews1975comparison, boughorbel2017optimal} to compare what variables show strong evidence to impact a participant's ability to guess the actual state of the video correctly. MCC is typically used for classification models to observe the classifier's performance. Here we treat human participant subgroups as classifiers and measure their performance with MCC. MCC takes the participant subgroup's guesses and the actual answers and breaks them up into the following categories: number of true positives (TP), number of true negatives (TN), number of false positives (FP), and number of false negatives (FN). The MCC metric ranges from -1 to 1, where 1 indicates total agreement between participant guess about the video and the actual state of the video, -1 indicates complete disagreement between participant guess about the video and the actual state of the video, and 0 indicates something similar to a random guess. To calculate the MCC metric for our human classifiers, we then use the following formula:

\begin{equation}\frac{(TP \times TN) - (FP  \times FN)}{\sqrt{(TP+FP)(TP+FN)(TN+FP)(TN+FN)}}\end{equation}

MCC is considered a more balanced statistical measure than an F1, precision, or recall score because it is symmetric, meaning no class (e.g., TP, TN, FP, FN) is more important than another.  

To compare MCC scores, we bootstrap samples from pairs of confusion matrices and compare their MCC scores. This process generates 10,000 bootstrapped samples of differences in correlation coefficients. We then compare the null hypothesis (difference equal to zero) to the bootstrapped distribution to measure the evidence level of biases and get a credibility interval on their strength. 

\textit{Logistic Regression:} To understand the relationship between matching demographics and guess accuracy we run a Bayesian logistic regression on matching demographics (age matches, gender matches, race matches). Logistic regression is a statistical analysis method used to model and predict binary outcomes (the participant's accuracy). Accuracy is equal to 1 if the participant's guess about the video was correct and 0 if it was incorrect. It utilizes prior observations from a dataset to establish relationships and make predictions based on specific variables. 

\textit{Accuracy Rate:} The performance metric we use to measure participant accuracy is the ratio of the correct guesses to the entire pool of guesses. The accuracy is thus equal to the sum of true positives and true negatives over the total number of guesses.

\section*{Data Availability}
\label{section:code}

Our full survey questionnaire, code, data, and codebook can be found on our GitHub repository. 

\href{https://github.com/juniperlovato/Diverse_Misinformation_Paper}{https://github.com/juniperlovato/DiverseMisinformationPaper}

Due to the nature of this research, participants of this study did not consent for their personally identifiable data to be shared publicly, so the full survey's raw individual level supporting data is not available. Aggregated and anonymized data needed for analysis can be found in our repository. 

\section*{Acknowledgements}

Institutional Review Board Approval: The survey in this project is CHRBSS (Behavioral) STUDY00001786, approved by the University of Vermont I.R.B. on 12/6/2021. The authors would like to thank Anne Marie Stupinski, Nana Nimako, Austin Block, and Alex Friedrichsen for their feedback on early drafts and Jean-Gabriel Young and Maria Sckolnick for comments on our analysis. The authors would also like to thank Engin Kirda and Wil Robertson for their contributions to an early survey prototype. This work is supported by the Alfred P. Sloan Foundation, The UVM OCEAN Project, and MassMutual under the MassMutual Center of Excellence in Complex Systems and Data Science. Any opinions, findings, conclusions, or recommendations expressed in this material are those of the author(s) and do not necessarily reflect the views of the aforementioned financial supporters.

\section*{Author contributions statement}

Author contributions: Conceptual: J.L., J.O., R.H., L.H-D.; Survey Development: J.L., J.O., R.H.; Survey Implementation: J.L., I.U.H., J.S-O., S.P.R., G.S.L.; Wrangling and Analysis: J.L., J.S-O., G.S.L., S.P.R., L.H-D.; Mathematical Model:  L.H-D., J.L.; All authors drafted the manuscript, revised it critically for important intellectual content, gave final approval of the completed version, contributed to the conception of the work, and are accountable for all aspects of the work in ensuring that questions related to the accuracy or integrity of any part of the work are appropriately investigated and resolved.

\section*{Competing interests statement}

The authors declare no Competing Financial Interests but the following Competing Non-Financial Interests: the author, Laurent H\'ebert-Dufresne, is the Editor-in-Chief for npj Complexity and was not involved in the journal's review of, or decisions related to, this manuscript.  

\bibliography{sample}

\end{document}